\documentstyle[12pt, epsf, epsfig]{article}
\begin{document}
\def\be{\begin {equation}}
\def\ee{\end {equation}}
\begin{center}
{\large \bf Perspectives for Direct CP Violation Tests \\
in Hyperon Decays \\}
\vspace {5 mm}
M. Conte$^1$, E. Di Salvo$^1$, A. Penzo$^2$ and M. Pusterla$^3$. \\
\vspace {0.5 cm}
{\small \it
(1) Dipartimento di Fisica dell'Universit\`a di Genova, INFN Sezione di
Genova, \\ Via Dodecaneso 33, 16146 Genova, Italy. \\
(2) INFN Sezione di Trieste, Via Valerio 2, 34127 Trieste, Italy. \\
(3) Dipartimento di Fisica dell'Universit\`a di Padova,
INFN Sezione di Padova, \\ Via Marzolo 8, 35131 Padova, Italy.}
\end {center}
\vspace {1.0 cm}
\centerline {ABSTRACT}
\vspace {0.5 cm}
{\small We discuss the perspectives of measuring 
direct CP violation parameters in hyperon decays 
by measurements with proton and/or antiproton
beams, to be carried out at various facilities, and in particular we
propose new experiments at the Tevatron and at the LHC.}
\vspace {1.5 cm}
\section {Introduction}

$~~~~$ Violation of CP symmetry is a crucial problem in 
particle physics, and has  attracted much theoretical and 
experimental interest, as witnessed by numerous review 
papers on the subject (see, for example, \cite{bigi,kay,he,bur}).
Up to now only one process has exhibited such a violation: 
$K^0_L$ decay into two pions. 
Various experimental tests have been proposed in
order to find further confirmations of such violations 
\cite{he} and, possibly, signatures of new physics 
\cite{bigi,kay,he,dan}. 

Some of these experiments, based on comparison between hyperon 
and antihyperon decays \cite{he,pak1,pak2}, were proposed more than 
ten years ago, but their practical realization appeared to be hard, 
mainly due to the difficulty in creating a sufficiently intense
antihyperon factory \cite{pak2}. 
Results from few pioneering experiments\cite{r6,ps,DCI} 
have a statistical precision too poor to be compared in a significant 
way with theoretical predictions.
More recently, \cite{pp} a viable scheme for a high precision 
experiment was proposed at FNAL, where the approved experiment 
was performed during the last fixed-target run, and 
represents a first serious step in the field of CP violation in the 
hyperon sector.

Meanwhile strong efforts are being engaged in a new generation of 
precision measurements, deeply exploring CP violation in the neutral 
kaon sector.
These will be followed by experiments in the beauty sector, both at 
hadron machines\cite{HERAB} and at $e^{+}-e ^{-}$ 
beauty factories \cite{bab}, that appear best suited for establishing 
indirect CP effects, similar to neutral kaons. 
    We consider here the perpectives for a full programme 
on direct CP violation tests in the hyperon sector, that could be 
performed by using $p{\bar{p}}$, $pp$ and ${\bar {p}}{\bar {p}}$ 
collisions at various facilities, including the Tevatron \cite{TeV} and the 
Large Hadron Collider (LHC) \cite{LHC}. 
These seem to be particularly favourable for testing
charge conjugation and CP invariance.
Our main purpose consists of illustrating the practical conditions required in 
this sense. Moreover we shortly review the problem of CP violation, especially 
in connection with (anti-)hyperon decays.

\section {Physics Outline}

$~~~~$ CP symmetry has attracted the interest of particle physicists since the
discovery of parity violation. In particular Landau \cite{la} pointed out that
parity violation should be
compensated by a charge conjugation asymmetry, in such a way that CP is 
conserved.
should leave the Lagrangian invariant.  On the other
hand a CP violation (or T violation, if CPT theorem is true, as we shall
assume from now on) is a much more serious and puzzling problem. Now we know
that the Landau hypothesis is likely to be true with a good approximation.
Indeed, in the very few cases when it has been measured, CP violation appears 
very small; furthermore all existing models
predict, in various systems, even smaller effects.

    We expect CP to be violated by weak interactions.
However, recently people have focused their interests also on strong 
interactions 
\cite{ZHua}, where CP could be violated by nontrivial configurations of
QCD theory \cite{pec}. These violations have not been observed; consequently
the theoretical question is still open \cite{pq}.

    As regards CP violations in weak interactions, the most popular model
is Kobayashi-Maskawa's \cite{km} (KM). They show that the weak Hamiltonian
matrix of quarks with three different quark families can carry a phase
which is not invariant by time reversal. To be more specific, let us consider
the contribution of CP violations in Feynman diagrams. They are

~i) ``penguin'' diagrams, exhibiting the ``direct'' CP violations due to the
decay process;

ii) ``box'' diagrams, corresponding to ``indirect'' CP violations; indeed in
some neutral mesons carrying
strangeness, charm or beauty (such as $K^0$, $B^0$, $D^0$), a CP violating mass
matrix can occur; these diagrams produce
typical particle-antiparticle mixings with oscillations of quantum numbers.

    In both cases one has interference among virtual exchanges of 
($u$,$c$,$t$)-quarks or ($d$,$s$,$b$)-quarks, so that the 
amplitude may be sensitive to the CP-noninvariant phase of the weak 
Hamiltonian. For the $K^0_L$ decay into two pions the main source of 
CP violation is the indirect one, characterized by the well-known 
parameter $\epsilon$, whereas the direct CP violation parameter 
$\epsilon'$ is affected by a large uncertainty and consistent with 
zero \cite{pdg,bigi}.

    Both effects can be looked for in different decay processes. Beauty
factories \cite{bab} appear to be the most promising for establishing 
indirect effects, similar to neutral kaons. 
However also direct effects are important, not only for further 
confirmations of CP violations, but also for a possible
discrimination among various models, which predict quite 
different values of violation parameters. Indeed, they vanish according to
the ``superweak'' Hamiltonian by Wolfenstein \cite{wo}, but not in
the KM model, because of ``penguin'' diagrams. To this end hyperon decays are 
very suitable, since they cannot exhibit indirect CP violations.

  A necessary condition for observing direct CP violations is that 
the process should consist of two interfering amplitudes at least, 
which differ in (strongly) conserved quantum numbers, e.g., 
isospin. In this connection Brown et al. \cite{pak1} point 
out that similar violations may be observed in processes 
of the type

\be
\Lambda \rightarrow p \pi^- (n \pi^0), \ ~~ \
\Sigma^+ \rightarrow n \pi^+ (p \pi^0),
\ee
but not in decays like \cite{pak2}

\be
\Sigma^- \rightarrow n \pi^-, \ ~~ \ \Xi^{0(-)} \rightarrow \Lambda
\pi^{0(-)},
\label{is}
\ee
where the final state is a pure isospin state. This argument might
be not completely correct, since in the KM model penguin diagrams 
are also involved in the decays (\ref{is}), which therefore would 
proceed through intermediate states of different isospin.

    The CP violation effects, which are expected to be very small
(although one predicts them to be quite different, according to the model and 
to the asymmetry 
parameter considered \cite{pak1,pak2}), would produce slight differences 
in hyperon - antihyperon decay width ($\Gamma$) and in the 
parameters $\alpha$, $\beta$ and $\gamma$, defined below.
These tiny differences look very hard to be detected from separate measurements
of the parameters. Only direct comparison
between the hyperon and the corresponding antihyperon decays,
measured simultaneously in the same apparatus, can hopefully
show evidence of such effects.
The decays we consider are of the type $Y \rightarrow B \pi$,
where $B$ is the final baryon. To this end some asymmetry parameters 
have been defined \cite{pak1,pak2}:

\be
A_1 = \frac{\Gamma - {\overline \Gamma}}{\Gamma + {\overline \Gamma}}, \ ~~ \
A_2 = \frac{\alpha + {\overline \alpha}}{\alpha - {\overline \alpha}}, \ ~~ \
A_3 = \frac{\beta + {\overline \beta}}{\beta - {\overline \beta}},
\ee
where bars refer to antiparticles. In particular $\Gamma$ is the
partial decay width (e.g., $\Lambda \rightarrow p \pi^-$), while

\be
\alpha (\beta) = \frac{2 Re (Im) [S^* P]}{|S^2|+|P^2|}
\ee
and  $S$ ($P$) is the s(p)-wave amplitude in the hyperon decay. It is worth
noticing that, while the CPT theorem implies equal total decay widths for
particle and antiparticle, CP violation allows differences between partial
decay widths.

   The decay distribution of the daughter spin ${1\over 2}$ baryon in the rest
frame of the parent hyperon is given by

\be {dP\over d\Omega} = {1\over 4\pi}(1+\alpha{\vec P_p} \cdot {\hat {p_d}})
\label {prob}, \ee
where $\vec P_p$ is the parent hyperon polarization and ${\hat {p_d}}$ is the
daughter baryon momentum direction in the rest frame of the parent. The
daughter itself is polarized with a polarization given by

\be \vec P_d = {(\alpha+{\vec P_p\cdot {\hat {p_d}}}){\hat {p_d}} +
\beta(\vec P_p \times {\hat {p_d}})+
\gamma({\hat {p_d}}\times(\vec P_p\times{\hat {p_d}}))\over
1+\alpha \vec P_p \cdot {\hat {p_d}}} \label {pol}, \ee
where $\alpha^2+\beta^2+\gamma^2=1$. We note that in the case of 
unpolarized parent the daughter is in a helicity state with a polarization 
$\alpha$.
Hence $\alpha$ and $\beta$ can be extracted from
polarization measurements of the hyperon and of the final baryon 
\cite{mrr}.
In this connection we point out that, while $\alpha$ can be 
measured from the decay angular distribution of an unpolarized
hyperon, $\beta$ cannnot be determined if the hyperon is 
unpolarized. However it is well known (see \cite{Heller} and 
references quoted therein) that most of the hyperons (and some 
antihyperons) are produced polarized transverse to their 
production plane.

\section {Preliminary Experimental Results}

$~~~~$ The existing experimental results on CP violation in hyperon decays 
are scarce and have a limited statistical precision; at the moment only 
three experiments give limits on $A_2$ for $\Lambda^0 
({\overline{\Lambda}}^0)$.
%\begin(itemize}

Experiment R608 at the ISR \cite{r6} studied the production and decay of
$\Lambda (\overline{\Lambda})$ from the
$pp \rightarrow {\Lambda} X$ 
$({\bar {p}}p \rightarrow {\overline{\Lambda}} X)$ reaction; the
measured decay asymmetries of the $\Lambda (\overline{\Lambda})$ give a ratio 
$(\alpha$ $P_{\Lambda})$ / $(\overline {\alpha}$ 
$P_{\overline{\Lambda}})$ = $-1.04\pm 0.29$; 
assuming the polarizations $P_{\Lambda}$ = $P_{\overline{\Lambda}}$, one
obtains $A_{2}$ = $-0.02\pm 0.14$.

Experiment PS185 at LEAR \cite{ps} studied the exclusive reaction  
${\bar {p}p} \rightarrow {\Lambda {\overline {\Lambda}}}$  
at 1.5 GeV/$\it c$ $\bar {p}$ incident momentum; in this case 
$P_{\Lambda}$ = $P_{\overline{\Lambda}}$ by C-parity 
conservation in strong interactions; it results an average value
$<A_{2}> = -0.07\pm 0.09$.

The third result comes from 
$e^{+} e^{-}$ $\rightarrow$ $J/\Psi$ $\rightarrow$ 
$\Lambda$ $\overline{\Lambda}$,
measured by DM2 at DCI-Orsay \cite{DCI}, and provides
$A_{2} = 0.01 \pm 0.10$.

These experimental results are limited by statistical errors that should 
be significantly improved in order to be compared in a useful way with 
theoretical predictions. 

%\end{itemize}

   More recently, a proposal (P871) \cite{pp} for a search of CP violation
in the decay of $\Xi (\overline{\Xi})$ and $\Lambda (\overline{\Lambda})$ 
hyperons was approved at FNAL; the experiment E871 has been
taking data during the last fixed-target Tevatron run. The measurement
consists of producing (unpolarized) beams of $\Xi (\overline{\Xi})$
and detecting their decays into $\Lambda (\overline{\Lambda})$, the latter 
exhibiting a helicity polarization $\alpha_{\Xi}$ 
($\overline{\alpha}_{\overline \Xi}$).
The $p (\bar{p})$ distributions from
the (polarized) $\Lambda (\overline{\Lambda})$ 
from  $\Xi (\overline{\Xi})$ decays 
are determined by the products
$\alpha_{\Xi}$ $\alpha_{\Lambda}$ 
($\overline{\alpha}_{\Xi}$ $\overline{\alpha}_{\Lambda}$);
the differences of these distributions will 
measure the sum ${A_2}^{\Xi} + {A_2}^{\Lambda}$ 
to a good accuracy. A statistically significant nonvanishing result
would represent a first serious signal of CP-violation in the hyperon sector.

\section {Hyperon Processes for CP-Violation Studies}

The study of CP-violation in the hyperon sector has so far
been considered for exclusive (2-body) reactions at low or intermediate energy,
for instance, hyperon-antihyperon production in $\bar{p}-p$ interactions,
and inclusive processes at very high energy, where intense hyperon
(antihyperon) beams can be produced. Now we propose new experiments where 
direct CP violations in hyperon decays could be observed, considering 
inclusive and exclusive reactions feasible at various facilities. 

\subsection{ Inclusive processes at high energy}

$~~~~$ The inclusive reactions we refer to are

~~i)  $p{\bar {p}} ~~ \rightarrow ~~ Y{\bar {Y}}X$  (feasible at Tevatron),

~ii)  $pp ~~ \rightarrow ~~ Y({\bar {Y}})X$ (feasible at LHC).

iii)  ${\bar {p}}{\bar {p}} ~~ \rightarrow ~~ {\bar {Y}}(Y)X$ (feasible at 
LHC).

In cases i) and ii) one assumes the hyperons and antihyperons to be
polarized (away from the forward direction), however in reaction ii) we 
expect a much lower antihyperon production cross section to test 
CP violation \cite{Heller}.

To this end ${\bar {p}}{\bar {p}}$ collisions have definite advantages.
More precisely, the production of hyperons in $pp$ and of
antihyperon in ${\bar{p}}{\bar{p}}$ collisions, should be equal, assuming
CP symmetry in strong interactions, and therefore the comparison of
decay distributions in these two cases becomes a direct test for CP violation.
This makes it possible to reformulate the tests proposed by Donoghue and
Pakvasa \cite{pak2}: in particular, according to their evaluations, the
largest signal could be found in $A_3$ ($10^{-3}$ to $10^{-2}$
in $\Xi^0 \rightarrow \Lambda \pi^0$, according to different models).

    Finally the suggested strategy of employing ${\bar {p}}$
beams could avoid the drawbacks connected with particular features of the
detector, clearly illustrated in ref. \cite{CIJ}.

    As far as the study of the inclusive reaction i) is concerned, we
obviously suppose to use the upgraded Tevatron at FNAL, where presumably the 
luminosity reaches the value
\be L_{p{\bar {p}}} = 1.61 \times 10^{32}~ cm^{-2}s^{-1}. \label {LTeV}\ee

\subsection{A possible ${\bar {p}}{\bar {p}}$ option for LHC} 

 In order to have ${\bar {p}}{\bar {p}}$ collisions, the
planned Large Hadron Collider (LHC) \cite{LHC} appears to be the most
suitable machine since consisting of a double ring with an antiproton
source nearby. Bearing in mind that the luminosity $L$ fulfils \cite{Sands}
the relation
\be L  \propto {N^2 \over b}, \label {Lgen} \ee
where $N$ is the total number of particles
per beam and $b$ is the number of interacting bunches, the
design parameters of LHC in its normal $pp$ mode, i. e.,
\be n_p = 1.05 \times 10^{11} ~~~ {\mathrm {protons~per~bunch,}}
\label {Nppb} \ee
\be b_p = 2835 ~~~ {\mathrm {proton~bunches,}} \label {nb1} \ee
\be N_p = b_p n_p = 2.98 \times 10^{14} \simeq 3 \times 10^{14} ~~~
{\mathrm {protons~per~beam,}} \label {Np} \ee
yield a luminosity
\be L_{pp} \simeq 10^{34} ~ cm^{-2}s^{-1}. \label {Lp} \ee

If we assume the $\bar{p}$ production rate to be the same as
the ACOL \cite{ACOL} yield of $N_{\bar{p}}=1.8 \times 10^{12}$
$\bar{p}$ 's per day, we plan to achieve
\be N_{\bar{p}} \simeq 10^{12} ~ \bar{p}/{\mathrm {beam.}} \label {Npbar} \ee
Then if the number of antiprotons per bunch equals the number of protons per 
bunch (see formula (\ref{Nppb})), we find
\be b_{\bar{p}} = 10 ~~~ {\mathrm {antiproton~bunches.}}
\label {nb2} \ee
Therefore, combining eqs. (\ref{nb1}),(\ref{Np}),(\ref{Lp}),(\ref{nb2}) and
(\ref{Npbar}), we obtain
\be L_{{\bar{p}}{\bar{p}}} = {b_p\over b_{\bar{p}}}
\left(N_{\bar{p}}\over N_p\right)^2 L_{pp} ~~ \simeq ~~ 3.15 \times 10^{31}
~ cm^{-2}s^{-1}, \label {Lpbar} \ee

which would allow us to obtain quite a good rate of events.

\subsection{Exclusive processes with $\bar{p}$ beams}

The process ${\bar {p}p} ~~ \rightarrow ~~ Y{\bar {Y}}$,  
feasible at FNAL $\bar{p}$ accumulator,
has the following advantages:
\begin{itemize}
\item { high luminosity;}
\item { clear kinematical configuration;}
\item { polarizations of Y and $\bar{Y}$ identical in binary reactions;}
\item { cascade decays of polarized hyperons providing access to $\beta$ parameters.}
\end{itemize}

The most promising scenario in this case is the reaction 
${\bar {p}p} \rightarrow \Xi{\bar {\Xi}}$, followed by the decays
$\Xi \rightarrow \Lambda \pi, (\bar {\Xi} \rightarrow \bar{\Lambda} \pi)$.
This reaction can be produced in interactions of the accumulated $\bar{p}$
beam with a jet target. The produced  $\Xi ({\bar {\Xi}})$ hyperons will 
have a polarization $P_Y$($\theta$*)  as a function of the production angle 
(vanishing at zero degrees) and therefore the polarization of the decay 
$\Lambda (\bar{\Lambda})$ will depend only on $\alpha_{\Xi}$
for forward production and on both $\alpha_{\Xi}$ and $\beta_{\Xi}$ for
production at an angle $\theta$*. We notice that it is not necessary to measure the
 $\Xi ({\bar {\Xi}})$ polarization, because, being the same for both, it cancels out
in the ratios measuring CP-violation effects.
In this way the final $p(\bar{p})$ distributions for fully reconstructed events
${\bar {p}p} \rightarrow 
\Xi(\rightarrow \Lambda \pi){\bar {\Xi}}(\rightarrow \bar{\Lambda} \pi)$
will determine a set of combinations  $\alpha_{\Xi}$$\alpha_{\Lambda}$ and
 $\beta_{\Xi}$ $\alpha_{\Lambda}$ for hyperons and antihyperons, in exactly the
same conditions, with minimal systematic biases.

\section{Conclusions} 

Since the discovery of CP violation several generations of experiments have 
been performed in the neutral kaon sector and others are planned in the beauty 
sector. In both cases the accessible CP effects are essentially "indirect". 

On the other hand, concerning "direct" CP violations, a full experimental 
program is still missing, although few exploratory measurements have been 
attempted.

We have outlined suggestions for such a program in the hyperon sector, on the 
basis of present and future facilities, among which the Tevatron and 
the LHC, using $p{\bar{p}}$, $pp$ and ${\bar {p}}{\bar {p}}$ 
collisions. 

\vspace {1.5 cm}

\begin {thebibliography} {}
%
\bibitem {bigi}
I. Bigi: Preprint UND-HEP-97-BIG 09.
\bibitem {kay}
B. Kayser: Preprint NSF-PT-97-4.
\bibitem {he}
Xiao-Gang He: Lecture given at the CCAST workshop on "CP violation and Various
Frontiers in Tau and Other Systems", Beijing, China, 11-14 August, 1997.
\bibitem {bur}
A.J.Buras: Preprint MPI-PhT/94-30, TUM-T31-64/94.
\bibitem {dan}
Dan-Di Wu: Preprint hep-ph/9710845.
\bibitem {pak1}
T.Brown, S.F.Tuan and S.Pakvasa: Phys. Rev. Lett. {\em 51} (1983) 1823.
\bibitem {pak2}
J.F. Donogue and S.Pakvasa: Phys Rev. Lett. {\em 55} (1985) 162.
\bibitem{r6}
P.Chauvat et al. (R608): Phys. Lett. {\em B163} (1985) 273. 
\bibitem{ps}
P.D.Barnes et al. (PS168): Phys. Lett. {\em B199} (1987) 147. 
\bibitem{DCI}
M.H.Tixier et al. (DM2): Phys. Lett. {\em B212} (1988) 523. 
\bibitem {pp}
Fermilab Proposal P-871, March 26, 1994.
\bibitem {HERAB}
H. Albrecht et al. (HERA-B): DESY-PRC 92/4 (1992).
\bibitem {bab}
BaBar Collaboration, Status report, SLAC-419 (1993);
BELLE Collaboration (KEKB), KEK-94-2 (1994). 
\bibitem {TeV}
Fermilab Luminosity Upgrade Group, Run II Handbook, Updated on February 4 
1998 (site: http://adwww.fnal.gov/lug/).
\bibitem {LHC}
The Large Hadron Collider, {\it The LHC Study Group}, CERN/AC/95-05(LHC).
\bibitem {la} 
L. Landau: Nucl. Phys. {\em 3} (1957) 254.
\bibitem {ZHua}
Z. Huang: Phys. Rev. {\em D48} (1993) 270.
\bibitem {pec}
R. Peccei: "CP violations", C. Jarlskog (ed.), World Sci. Singapore, 1989,
p. 503.
\bibitem {pq}
R. Peccei and H. Quinn: Phys. Rev. Lett. {\em 38} (1977) 1440;
Phys. Rev. {\em D 16} (1977) 1791.
\bibitem {km}
M. Kobayashi and T. Maskawa: Prog. theor. Phys. {\em 49} (1973) 652.
\bibitem {pdg}
Particle Data Group (R. Barnett et al.): Phys. Rev. {\em D54}, (1996) 1.
\bibitem {wo}
L. Wolfenstein: Phys. Rev. Lett. {\em 13} (1964) 562.
\bibitem {mrr}
R.E. Marshak, Riazuddin and C.P. Ryan: "Theory of Weak
Interactions in Particle Physics", Wiley, New York, 1969
\bibitem {Heller}
K. Heller: Proceedings "Polarization Dynamics in Particle and Nuclear
Physics", Y. Onel, N. Paver and A. Penzo (ed.), World Sci., Singapore, 
1995, p. 231. (to be corrected)
\bibitem {CIJ}
C.J.-C.In, G.L. Kane and P.J. Molde: Phys. Lett. {\em B317} (1993) 454.
\bibitem {Sands}
M. Sands: The Physics of Electron Storage Rings, SLAC Report 121 (1970).
\bibitem {ACOL}
Design Study of an Antiproton Collector for the Antiproton Accumulator (ACOL),
{\it Design Study Team}, CERN Yellow Report 83-10.
%
\end {thebibliography}
\end {document}